\documentclass[a4paper]{jpconf}
\usepackage{graphicx}
\usepackage{wrapfig}
\usepackage{amsmath}
\usepackage{amssymb}
\usepackage{color}
\usepackage[utf8]{inputenc}

\begin{document}
\title{Theory of high pressure hydrogen, made simple}
\author{Ioan B Magdău, Floris Balm and Graeme J Ackland}
\address{CSEC, SUPA, School of Physics and Astronomy, The University of Edinburgh, Edinburgh EH9 3JZ, United Kingdom}
\ead{i.b.magdau@sms.ed.ac.uk, gjackland@ed.ac.uk}

\begin{abstract}
Phase I of hydrogen has several peculiarities. Despite having a
close-packed crystal structure, it is less dense than either the low
temperature Phase II or the liquid phase. At high pressure, it
transforms into either phase III or IV, depending on the temperature.
Moreover, spectroscopy suggests that the quantum rotor behaviour
disappears with pressurisation, without any apparent phase transition \cite{howie2014phonon}.
Here we present a simple thermodynamic model for this behaviour based
on packing atoms and molecules and discuss the thermodynamics of
the phase boundaries. We also report first principles
molecular dynamics calculations for a more detailed look at the same phase transitions.
\end{abstract}

\section{Introduction}

In 1935 Wigner and Huntington proposed a metallic modification of
hydrogen as an atomic phase at high pressure. They calculated that
metallisation would occur at about tenfold compression, which is
remarkably close to modern estimates. They also quoted a transition
pressure of 25GPa, which has proved over an order of magnitude too
low, and has been widely ridiculed. This should stand as testament
that calculated volumes are more reliable than pressures.

The experimental search for crystalline metallic hydrogen continues,
and with numerous non-metallic phases reported, it has become clear that
a different physical picture from Wigner and Huntingdon's ``atomisation
begets metallisation'' is required.

The structure of solid hydrogen up to 250GPa is now generally agreed
upon. Phase I is a hexagonal close-packed molecular liquid. Here, the
high-school picture of H$_2$ as a dumbbell molecule is misleading,
because at low pressure H$_2$ behaves as a free molecular rotor, and
the $J=0$ quantum ground state is spherical. Hence, phase I can be
thought of as simply close packing of spherical objects. As
pressure is increased the molecules interact with each other and $J$
ceases to be a good quantum number. At low temperature this leads to
a ``broken symmetry'' Phase II, where the rotation has stopped. There
have been numerous predictions for the crystal structure of this
phase, which is normally
regarded as a structure near to hcp in which molecular orientation is
fixed. 
At high temperature, the melt
line shows a maximum around 900K/70GPa\cite{bonev2004quantum,eremets2009evidence,howie2015raman}; if pressure is increased
further the melting temperature drops, meaning the liquid is denser than the close-packed Phase I. 

At high pressures, where the mechanical
work of compression ($PV$) approaches the molecular binding energy, complex phases are observed.
Theory predicts that these are based around a new motif - weakly bound
molecules arranged into hexagonal trimers\cite{pickard2012density,magdau2013identification,liu2013proton}. Distinct molecules
are still observed, albeit with short-lived, weaker and longer bonds.
In the low temperature ``Phase III''
all molecules are in such trimers, however, at high temperature
``Phase IV'' appears to comprise alternating layers of trimers and
relatively freely rotating
molecules\cite{magdau2013identification, eremets2011conductive}.
If the proton motions
are treated quantum mechanically, either by path integral sampling or
through consideration of tunneling probabilities, the many-proton
wavefunction attains P$6/mmm$ symmetry directly and the weak molecules
manifest as a correlation in the proton wavefunction.

These phases can all be reproduced by considering classical protons
interacting via forces calculated in the density functional theory.
The calculated pressure and temperature phase boundaries depend on the
details of the calculation which typically involve trade-offs between
system size, exchange correlation treatment, pseudopotential fidelity, 
quantisation of nuclear motion and k-point sampling. 

It is widely assumed that quantised proton motion and advanced treatment of
exchange-correlation for van der Waals bonding are essential to
describe the bonding\cite{li2013classical}. So it is curious that
the ``quantum rotor'' phase I is well explained by molecular dynamics
using PBE and classical rotors\cite{bonev2004quantum}. 
Given that the melting point maximum was
predicted by classical molecular dynamics
simulation with remarkable accuracy, one has to
believe that if those quantum effects are essential, they are
exactly compensated by an incorrect piece of classical physics.

In this paper we examine what is the minimal physics required for an 
understanding of the hydrogen phase diagram. We do not aspire to obtain accurate pressures or temperatures, rather the questions to address are:

\begin{enumerate}

\item
Is there a simple explanation for the mixed atomic-molecular phase IV?
\item
Does phase I comprise free rotors at all pressures?
\item
How does the melt become denser than the close-packed solid phase I?
\end{enumerate}

\section{A simple packing model}
Despite the wide range of candidate structures reported from ab initio
structure-search techniques\cite{pickard2012density, pickard2007structure,geng2012high},
there are essentially three motifs that are sufficient to capture the physics of the phase diagram: rotating molecules, non-rotating molecules, and layers comprising
short-lived molecules where the time-averaged, indistinguishable-atom
positions form hexagonal layers. Structure search reveals numerous
possible small distortions which molecular dynamics or path integral
methods show are eliminated by elevated temperatures and/or quantum
behaviour of the protons\cite{liu2013proton, cogent,magduau2013identification}.
Our simple model for this considers three objects:
\begin{itemize}
\item S: Spherical molecules, 
which correspond to a $J=0$ quantum rotor
 ground state, or a time-averaged classical free rotor. S objects
 have high volume and high internal entropy. 

\item R: Rod-like molecules, corresponding to the non-rotating type,
 with length similar to the diameter of S. 
R molecules have smaller volume than S, and zero internal entropy or energy.

\item A: Spherical ``atoms'', 
about half that of S. They have no internal entropy, and high internal 
energy due to the broken bond.
\end{itemize}

\begin{wrapfigure}{rh}{0.45\textwidth}
\begin{center}
\vspace{-25pt}
\includegraphics[width=0.45\textwidth]{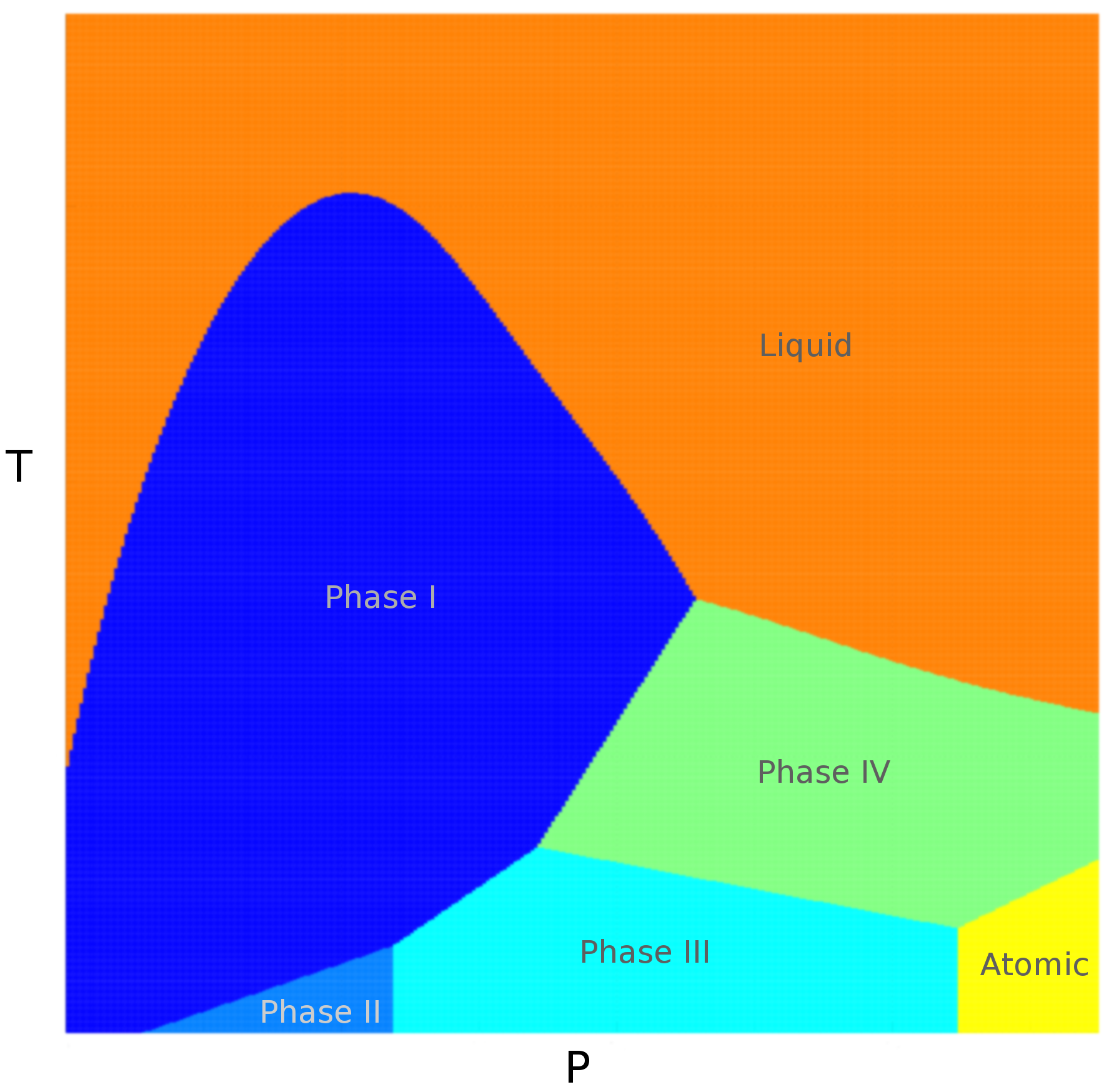}
\end{center}
\vspace{-20pt}
\caption{Example phase diagram from the simple packing model 
Code and parameters used are available in supplemental materials.}
 \vspace{-15pt}
\label{phased}
\end{wrapfigure}

In our model, phase I comprises S-type objects, phase II R-type, phase
III R-type (better packed, but elongated), phase IV is a binary mixture of A and S. The liquid
contains S, R and A objects according to their Boltzmann weights. A
packing fraction and an internal energy is assigned to each
crystal structure. The total free energy of each phase is then the
sum of the energies of the crystal and its component objects. We
have considered other crystalline arrangements within the model, and
the only competitive structure is close-packing involving a single H
atom (labeled ``Atomic'' in figure \ref{phased}).

We can model the hydrogen molecule as two spherical atoms with centers
separated by about 0.74\AA. At atmospheric pressure, the
density is 23 cm$^3$/mol, (about 39 $\AA^3$ per molecule) indicating
that the molecules are far apart compared to their bond length. At low
pressure, hydrogen is extremely compressible, which implies that in
this regime, close-packing of rigid spheres is not a good approximation. 
The close-packed density for such S-objects, i.e. where the
intermolecular distance becomes close to double the bohr radius,
 is reached at around
200GPa\cite{eremets2011conductive, caillabet2011multiphase,mao1994ultrahigh}
which is where transitions to phase III and IV start to happen.

We have no explicit treatment of compressibility within a phase. This
is because to calculate phase boundaries using the geometric model
presented here requires only that the volume difference between
competing phases is similar along the phase boundary (c/f figure
\ref{compar}).  This breaks down for low pressures where an additional
parameter is required (see supplemental).

The hcp
structure has a packing fraction of 0.7405, which is maintained by an
affine deformation to nonideal $c/a$ by transforming spheres into oblate
spheroids.  Naively, one might expect a transformation directly from
close-packed S to A, at a pressure which favours the smaller objects,
however, a mixed S-A structure can have denser packing provides it
adopts the, MgB$_2$ structure, P$6/mmm$, one of only two structures
which has a lower free energy than hcp for binary hard spheres\cite{LSMC}.
The time-averaged structure of SA$_2$ Phase IV is precisely
this\cite{magdau2013identification, liu2013proton}. 

Therefore the pressure stability of this apparently complex Phase IV
can be attributed to efficient SA$_2$ packing.  The temperature
stabilisation of Phase IV vs III comes from the entropy of S objects.
In molecular dynamics, it is observed that the atoms in phase IV form
short-lived molecules, like our R-objects, so we assign a lower energy
to IV the atomic close-packing phase objects.  

The melting point maximum requires that the liquid has higher
compressibility than the solid.  In the model, liquid contains all
three objects weighted by their free energy, so the increased population
of smaller type R and A objects causes the liquid to
become denser at high pressure.

The thermodynamic explanation for the phase transitions in this picture is:
\begin{itemize}
\item I $\rightarrow$ II.  Quadrupole-quadrupole interaction favours
  phase II. Pressure favours phase II due to R being smaller than S,
  despite inefficient packing of phase II to minimise EQQ.  S-object entropy
  favours phase I.
\item I $\rightarrow$ IV. Pressure favours phase IV's denser AB$_2$
  packing. S-object entropy favours phase I.
\item I $\rightarrow$ liquid. Entropy favours the liquid phase.
  Liquid has less efficient packing, but contains smaller objects,
  especially at high pressure.  So pressure initially favours I, but
  subsequently favours the liquid.

\item III$\rightarrow$IV. Entropy gain from S-objects and volume
  decrease due to efficient AB$_2$ packing.
\end{itemize}

PIMD calculations have shown that zero point motion reduces all
transition temperatures, so nuclear quantum effects can be included as
an offset of the temperature axis. The phase diagram corresponding to
this model is shown in figure \ref{phased}.

\section{A classical-proton AIMD model for the free and arrested rotor phases}
We have carried out extensive ab initio molecular dynamics simulations
on the phase boundaries around Phase I using the CASTEP code with PBE
exchange and classical protons. We carry out NPT simulations at
100GPa, starting at 0K in P$6_3/m$. 
NPT enables us to determine the c/a
ratio of hexagonal phases, and the volume changes on transition. The
rapid heating rates and finite system size mean that the transition
temperatures will be an overestimate of their 

\begin{wrapfigure}{lh}{0.45\textwidth}
\begin{center}
\vspace{-30pt}
\includegraphics[width=0.45\textwidth]{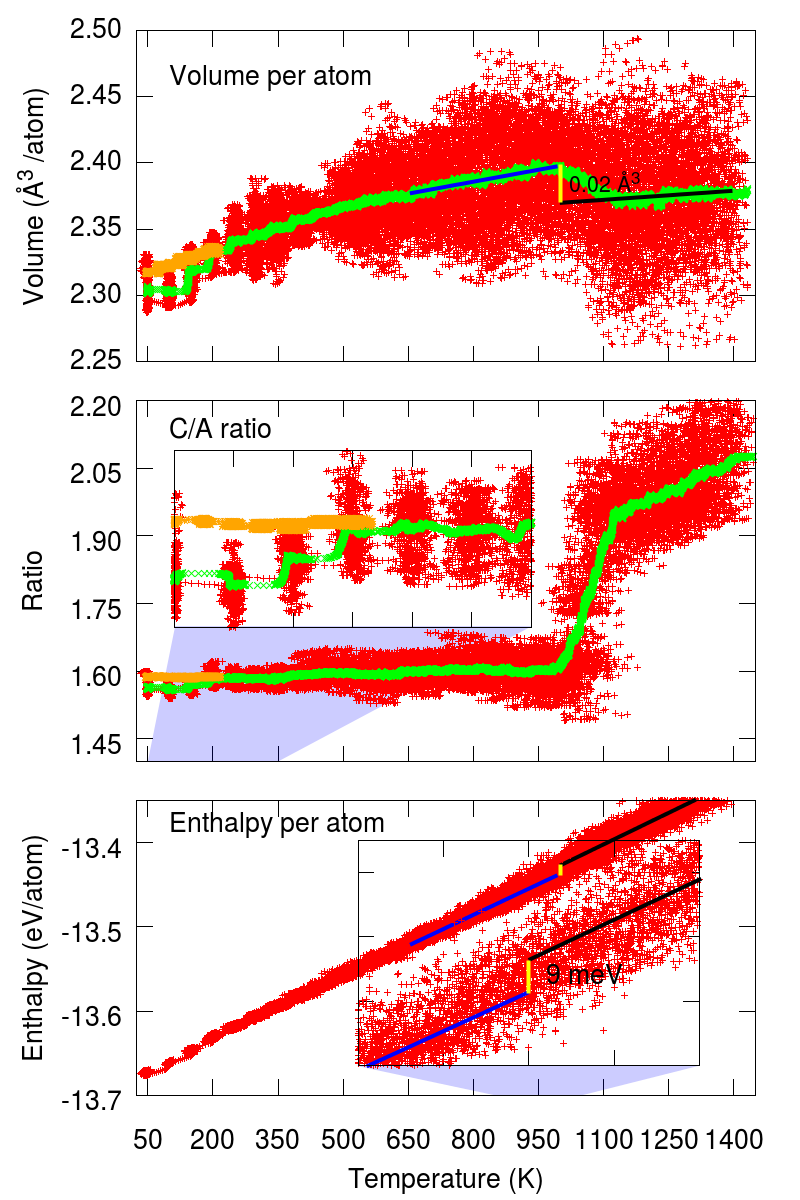}
\end{center}
\vspace{-20pt}
\caption{Volume, c/a and enthalpy from a 13.5ps NPT simulation with 288 atoms,
  0.5fs timestep, $\Gamma$ sampling at 100GPa. Thermostat
  temperature  was increased by 50K every 0.5ps.
Red crosses depict instantaneous values, light green lines 
are a 100fs window running
 average. Orange lines are a running average
 over a similar simulation with (4x4x4) k-point sampling heated at
 25K/150fs. The yellow lines are estimates for the enthalpy and
 volume drops at melting, obtained from linear regression fits just
 below and above the melting point.}
 \vspace{-0pt}
\label{compar}
\end{wrapfigure}
\noindent
 true    thermodynamic
values, nevertheless, the phase transitions  observed at this
level of theory are in qualitative accordance with experiment.

\begin{wrapfigure}{rh}{0.65\textwidth}
\vspace{-50pt}
\includegraphics[width=0.65\textwidth]{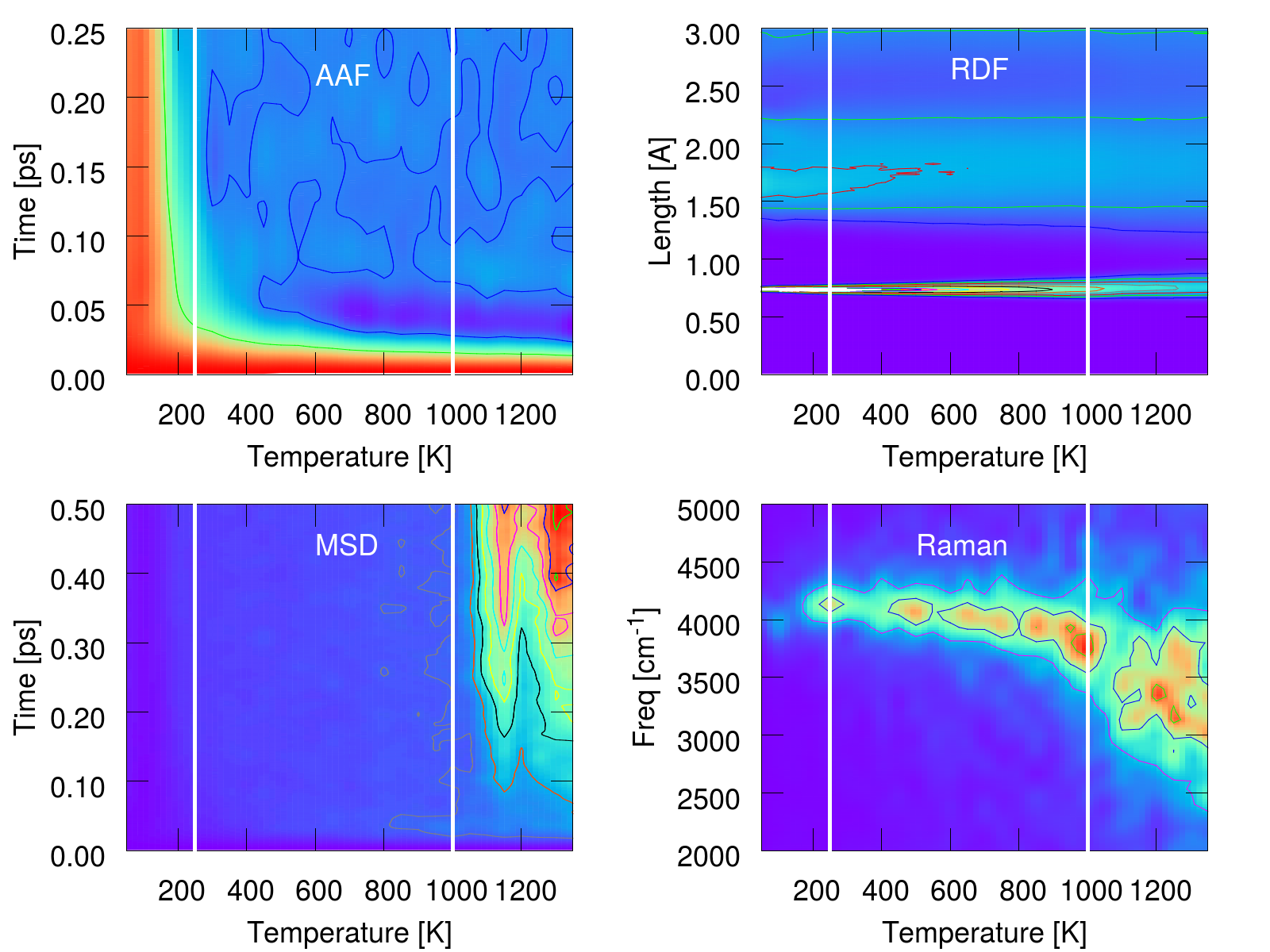}
\vspace{-25pt}
\caption{
Analysis of heating MD runs at 100GPa (a)
AAF vs temperature. Below 150K, in phase II, there is no rotation. 
Above 150K,  in phase I, AAF vanishes within 10 fs, but only at
at higher temperature, is rotation through 90 degrees free enough 
to see anticorrelation at 40fs.  (b) RDF: Phase II has
two molecular neighbours peaks at 1.7 and 2.1 \AA, which merge in phase I. 
all peaks broaden with T. (c) Mean squared displacement is low in
Phase II, indicating molecular libration. In Phase I, it is higher due to rotations. 
Above 1100K, a linear increase in MSD indicates melting. (d) Raman vibron frequency 
obtained via projection\cite{ackland2013efficacious},  dropping with increasing T, and further at the melting
transition. This is consistent with experimental data\cite{howie2015raman}.
 }
 \vspace{-35pt}
\label{heatAAF}
\end{wrapfigure}

On heating at 100GPa, (figure \ref{compar}) we observe the 
 II - I
transformation at 100K,
and melting at 1100K. The close-packed phase I has the lowest
density of the three. The phase transitions are evident in movie
visualisation. II$\rightarrow $I is most clearly shown in the
angular autocorrelation function (AAF, figure \ref{heatAAF}a), while
melting becomes evident in the mean squared displacement (MSD figure
\ref{heatAAF}c).

 Direct measurement of quadrupole interaction energy from the
 electronic structure is swamped by fluctuations in the MD, but we can
 measure it via orientational correlations. 
The quantity EQQ
 (figure \ref{eqq}) is the interaction energy of two linear quadrupoles
 pointing in directions, $k$, $k'$, separated by a vector $r$. The molecular quadrupole moment
for hydrogen is around $Q=0.26D\AA$ \cite{orcutt1963influence}. and the
 ensemble average of EQQ is the quadrupole-quadrupole energy.
{\small
\begin{equation*}
\begin{split}
<EQQ> = <(3Q^2/4\pi\epsilon_0r^5)[35(k.r)^2(k'.r)^2-5(k.r)^2\\
-5(k'.r)^2+2(k.k')^2-20(k'.r)(k.r)(k.k')+1]>
\end{split}
\end{equation*}
} A series of simulations of pressurisation at 600K (Fig \ref{Press})
showed phase I persisting until about 200GPa at which point it
transformed into a disordered structure.  The enhanced AAF, indicating
the arrest of the rotors, shows this is not a liquid.  The
disappearance of the strong vibron suggests the structure is similar
to the chain-like structure reported elsewhere\cite{thisvolume}.

AAF, RDF, MSD graphs confirm the broken symmetry - free rotor - liquid sequence
with temperature and the weakening of the vibron with increased pressure,
followed by extreme broadening and further softening, plus suppression of
molecular rotation in Phase III.

\subsection{k-point convergence}
Previous work has reported P$ca2_1$,
P$6_3m$\cite{pickard2007structure} and P$2_1c$\cite{li2013classical}
as the most stable structure for phase II.  From static relaxations of
MD snapshots, we found this stability sensitive to $k$-point
sampling\cite{thisvolume}. Here, for phase II identified by EQQ,
$\Gamma$-only runs favour low $c/a$ ratio whereas at full $k$-point
convergence P$6_3/m$ remains stable until th eII-I transition. $k$-point sampling issues also appear
in the melting: sampling with a single $k$-point we observed melting
at 1100K; This calculation is essentially a repeat of previous
work\cite{bonev2004quantum}.  However, using a denser $k$-point mesh
systematically raised the melting point by several hundred degrees. 

It is unusual to require dense k-point sampling in molecular crystal, but
careful convergence is needed to properly resolve the quadrupole
moments in phase II, which are crucial to the ordering. Meanwhile, in
the liquid phase, free diffusion allows the atoms to arrange
themselves in a way most favourable for minimising the $\Gamma$-point
electron energy at the expense of unsampled contributions from other
parts of the Brillouin Zone\cite{thisvolume}.

\subsection{Lattice parameters}

Our simulation spontaneously transformed into the hcp structure for
phase I. We find the c/a ratio to be essentially independent of
temperature, but to decrease with increased pressure from ideal to
about 1.57 at 150GPa. This, and the PV equation of state, is in
good agreement with the experiment\cite{loubeyre1996x}.

\subsection{Local order}
\begin{figure}[h!]
\begin{minipage}{0.33\textwidth}
\includegraphics[width=\textwidth]{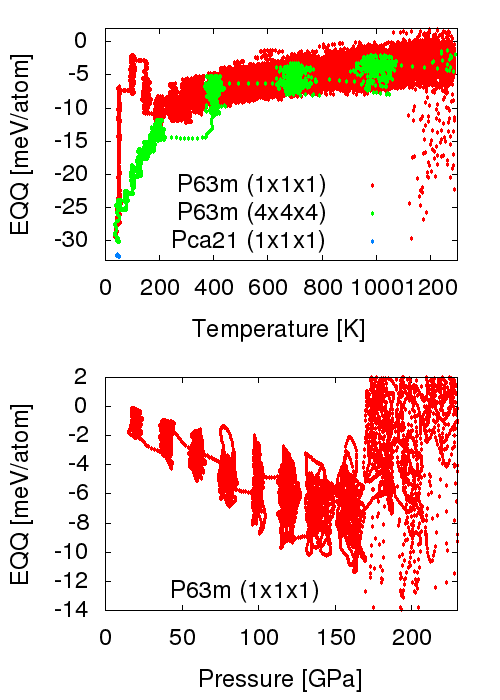}
\vspace{-25 pt}
\caption{EQQ for heating at 100GPa (top) and
 pressurising at 600K (bottom). Phase II is unstable with
 $\Gamma$-point sampling (red), but stabilised by slower heating and more $k$-point
 sampling (green). The blue dot shows EQQ for $Pca21$.}
\label{eqq}
\end{minipage}
\hspace{0.01\textwidth}
\vline
\hspace{0.01\textwidth}
\begin{minipage}{0.64\textwidth}
\includegraphics[width=\textwidth]{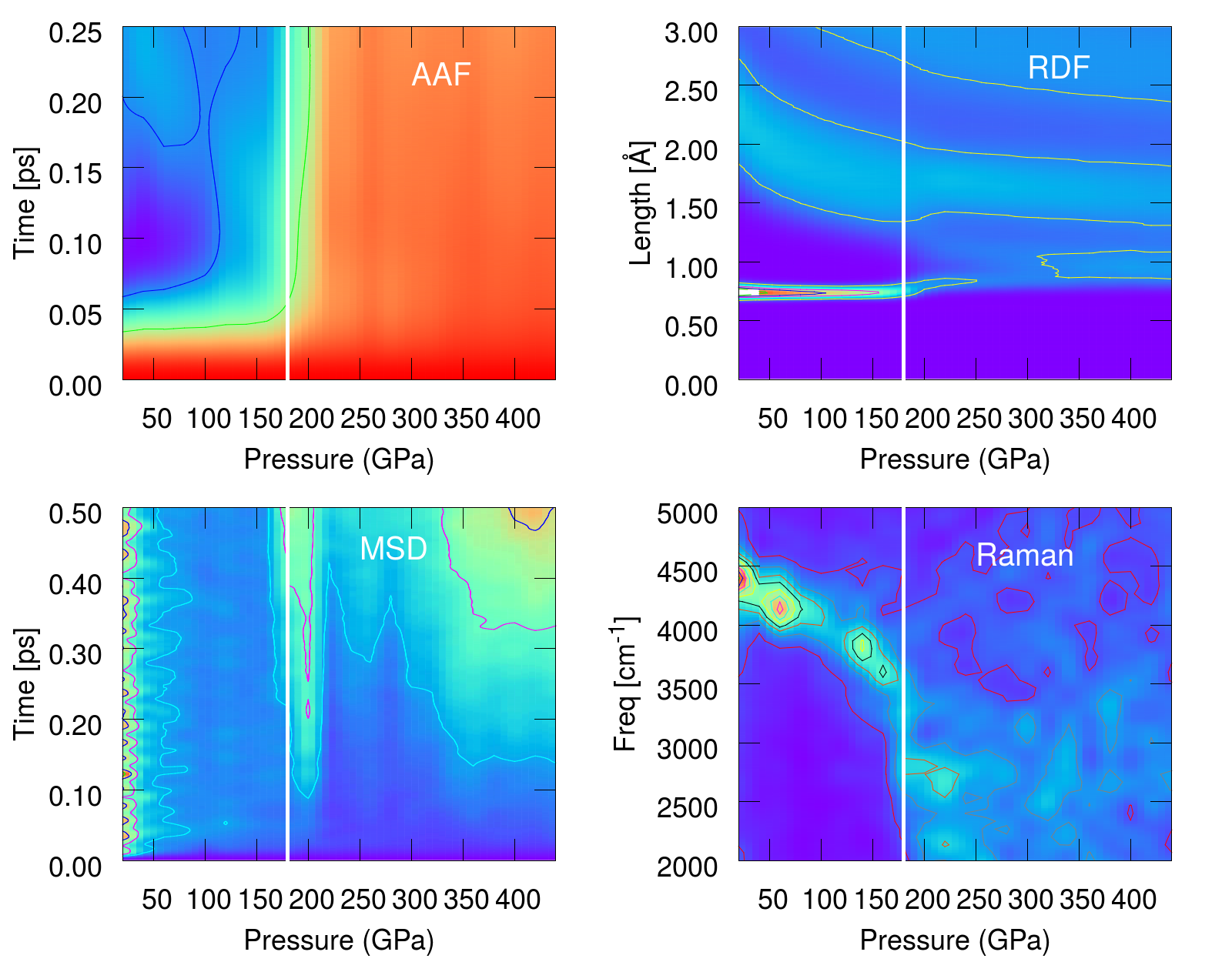}
\vspace{-20 pt}
\caption{ Analysis of pressure dependence at 600K: (a) AACF (b) RDF
 (c)MSD (d) Raman vs P. Two simulations were started from the
 equilibrated cell at 100GPa, 600K (phase I) and pressurised/depressurised 
 adiabatically by 20GPa every 0.5ps. The simulations were run with
 $\Gamma$ k-point and 288 atoms.}
\label{Press}
\end{minipage}
\vspace{-15pt}
\end{figure}

Phase I has no long range order, but the negative $\langle$EQQ$\rangle$
(Fig. \ref{eqq}) indicates short ranged order and
increasing coupling between the rotors with pressure of
 a few meV, increasing with pressure
approximately as $V^{-\frac{5}{3}}$. 
The correlation increases as the rotors are become coupled, leading to 
gradual breakdown in $J$
as a good quantum number. Despite the distinctive
signature in spectroscopy\cite{howie2014phonon}, this correlation does not
become long-ranged and so does not constitute a thermodynamic phase transition.

At low temperature, EQQ is sufficient to
cause long-range ordering in the broken-symmetry phase II,
with a binding of up to 30meV, 
enough to be the cause of its stabilisation below 100K.

\section{Conclusions}
We have presented a model which distills the essential physics
determining the phase stability in hydrogen. Phases II and III are
energetically favored, through quadrupole interaction and electron
delocalisation respectively. Phases I and IV are entropically
favoured, primarily due to the high entropy of rotating molecules,
which can be understood classically or quantum mechanically. The
J=0 quantum rotor is an intrinsically large object: the higher
compressibility of the liquid compared to phase I can be understood by
the increased population of ``smaller'' molecules in higher energy states.

The success of the classical MD models can be
attributed to this near equivalence of quantum uncertainty described
by the wavefunction, and classical uncertainty described by the
entropy.

The free rotors in phase I become increasingly correlated as pressure
increases, due to increased EQQ and steric
effect. The many-body wavefunction for this behaviour is not presented
here, but clearly the excited states observable in spectroscopy will
not have the characteristic overtones of a free rotor. This change in
behaviour does not indicate a phase transition.

\subsection*{Acknowledgements}
We  thank EPSRC for support, under UKCP grant K01465X and a
studentship for IBM. Additional data including the code for the simple
model and MD trajectories can be found at doi.???.????.

\section*{References}

\bibliographystyle{iopart-num}
\bibliography{Refs}

\end{document}